\documentclass[twocolumn,prl,a4paper,superscriptaddress,showpacs,
preprintnumbers,tightenlines,footnote]{revtex4}
\draft 
\usepackage{amsmath}
\usepackage{graphicx}
\usepackage{epsfig}

\usepackage{dcolumn}
\newcolumntype{B}{D{B}{}{-1}}

\begin{document}

\setlength{\unitlength}{1pt}

\title{ \quad\\[0.5cm] Measurements of exclusive {\boldmath $B_s^0$} decays
at the {\boldmath $\Upsilon$(5S)} resonance}

\affiliation{Budker Institute of Nuclear Physics, Novosibirsk}
\affiliation{Chiba University, Chiba}
\affiliation{University of Cincinnati, Cincinnati, Ohio 45221}
\affiliation{Department of Physics, Fu Jen Catholic University, Taipei}
\affiliation{The Graduate University for Advanced Studies, Hayama}
\affiliation{Hanyang University, Seoul}
\affiliation{University of Hawaii, Honolulu, Hawaii 96822}
\affiliation{High Energy Accelerator Research Organization (KEK), Tsukuba}
\affiliation{University of Illinois at Urbana-Champaign, Urbana, Illinois 61801}
\affiliation{Institute of High Energy Physics, Vienna}
\affiliation{Institute of High Energy Physics, Protvino}
\affiliation{Institute for Theoretical and Experimental Physics, Moscow}
\affiliation{J. Stefan Institute, Ljubljana}
\affiliation{Kanagawa University, Yokohama}
\affiliation{Korea University, Seoul}
\affiliation{Kyungpook National University, Taegu}
\affiliation{Swiss Federal Institute of Technology of Lausanne, EPFL, Lausanne}
\affiliation{University of Ljubljana, Ljubljana}
\affiliation{University of Maribor, Maribor}
\affiliation{University of Melbourne, School of Physics, Victoria 3010}
\affiliation{Nagoya University, Nagoya}
\affiliation{Nara Women's University, Nara}
\affiliation{National Central University, Chung-li}
\affiliation{National United University, Miao Li}
\affiliation{Department of Physics, National Taiwan University, Taipei}
\affiliation{H. Niewodniczanski Institute of Nuclear Physics, Krakow}
\affiliation{Nippon Dental University, Niigata}
\affiliation{Niigata University, Niigata}
\affiliation{University of Nova Gorica, Nova Gorica}
\affiliation{Osaka City University, Osaka}
\affiliation{Osaka University, Osaka}
\affiliation{Panjab University, Chandigarh}
\affiliation{Peking University, Beijing}
\affiliation{RIKEN BNL Research Center, Upton, New York 11973}
\affiliation{Saga University, Saga}
\affiliation{Seoul National University, Seoul}
\affiliation{Shinshu University, Nagano}
\affiliation{Sungkyunkwan University, Suwon}
\affiliation{University of Sydney, Sydney, New South Wales}
\affiliation{Tata Institute of Fundamental Research, Mumbai}
\affiliation{Toho University, Funabashi}
\affiliation{Tohoku Gakuin University, Tagajo}
\affiliation{Tohoku University, Sendai}
\affiliation{Department of Physics, University of Tokyo, Tokyo}
\affiliation{Tokyo Institute of Technology, Tokyo}
\affiliation{Tokyo Metropolitan University, Tokyo}
\affiliation{Virginia Polytechnic Institute and State University, Blacksburg, Virginia 24061}
\affiliation{Yonsei University, Seoul}
  \author{A.~Drutskoy}\affiliation{University of Cincinnati, Cincinnati, Ohio 45221} 
  \author{K.~Abe}\affiliation{Tohoku Gakuin University, Tagajo} 
  \author{I.~Adachi}\affiliation{High Energy Accelerator Research Organization (KEK), Tsukuba} 
  \author{H.~Aihara}\affiliation{Department of Physics, University of Tokyo, Tokyo} 
  \author{D.~Anipko}\affiliation{Budker Institute of Nuclear Physics, Novosibirsk} 
  \author{A.~M.~Bakich}\affiliation{University of Sydney, Sydney, New South Wales} 
  \author{E.~Barberio}\affiliation{University of Melbourne, School of Physics, Victoria 3010} 
  \author{I.~Bedny}\affiliation{Budker Institute of Nuclear Physics, Novosibirsk} 
  \author{U.~Bitenc}\affiliation{J. Stefan Institute, Ljubljana} 
  \author{I.~Bizjak}\affiliation{J. Stefan Institute, Ljubljana} 
% ...
\collaboration{The Belle Collaboration}

\date{\today}

\tighten

\begin{abstract}
Several exclusive $B_s^0$ decays are studied
using a 1.86\,fb$^{-1}$ data sample collected
at the $\Upsilon$(5S) resonance with the Belle
detector at the KEKB asymmetric energy $e^+ e^-$ collider.
In the $B_s^0 \to D_s^- \pi^+$ decay mode we find 
10 $B_s^0$ candidates and measure the corresponding branching fraction. 
Combining the $B_s^0 \to D_s^{(*)-} \pi^+$, $B_s^0 \to D_s^{(*)-} \rho^+$,
$B_s^0 \to J/\psi \phi$ and $B_s^0 \to J/\psi \eta$ decay modes,
a significant $B_s^0$ signal is observed.
The ratio $\sigma (e^+ e^- \to B_s^* \bar{B}_s^*) /
\sigma (e^+ e^- \to B_s^{(*)} \bar{B}_s^{(*)}) = (93 ^{+7}_{-9} \pm 1)\%$
is obtained at the $\Upsilon$(5S) energy, indicating that
$B_s^0$ meson production proceeds
predominantly through the creation of $B^*_s \bar{B}^*_s$ pairs.
The $B_s^0$ and $B_s^*$ meson masses are measured to be 
$M(B_s^0) = (5370 \pm 1 \pm 3)\,$MeV/$c^2$ and 
$M(B_s^*) = (5418 \pm 1 \pm 3)\,$MeV/$c^2$.
Upper limits on the
$B_s^0 \to \gamma \gamma$, $B_s^0 \to \phi \gamma$, $B_s^0 \to K^+ K^-$
and $B_s^0 \to D_s^{(*)+} D_s^{(*)-}$ branching fractions are also reported.
\end{abstract}


\maketitle

\tighten

{\renewcommand{\thefootnote}{\fnsymbol{footnote}}}
\setcounter{footnote}{0}

\section{Introduction}

A considerable $B_s^0$ production rate has been recently measured
in $e^+ e^-$ collisions at the energy of the 
$\Upsilon$(5S) resonance \cite{cleoi,bela}.
Thus, high luminosity $e^+e^-$ $B$-factories have 
great potential for studies of exclusive $B_s^0$ decays.
Although several $B_s^0$ decay channels have been recently observed 
by the Tevatron experiments \cite{teva,tevb},
a number of $B_s^0$ decay modes can be better measured at $e^+ e^-$ colliders
running at the $\Upsilon$(5S) energy. The detectors taking data
at the $\Upsilon$(5S) have many advantages in studies of $B_s^0$ decays,
such as high photon and $\pi^0$ reconstruction efficiency,
trigger efficiency of almost 100$\%$ for hadronic modes and
excellent charged kaon and pion identification.
The possibility of partial reconstruction of specific $B_s^0$ decays
and a model-independent determination of the number of initial $B_s^0$ mesons, 
which opens the possibility of precise absolute $B_s^0$ branching fraction
measurements, are additional advantages
of $B_s^0$ studies at $e^+ e^-$ colliders running
at the $\Upsilon$(5S).

In this paper we report measurements of exclusive $B_s^0$ decays
based on an $\Upsilon$(5S)
data sample of 1.86\,fb$^{-1}$, collected
with the Belle detector \cite{belle} at the KEKB asymmetric energy
$e^+ e^-$ collider \cite{kekb}.
This data sample is more than four times larger than
the 0.42~fb$^{-1}$ dataset collected at the $\Upsilon$(5S)
by the CLEO experiment in 2003 \cite{cleoe}, where first evidence
of exclusive $B_s^0$ decays at the $\Upsilon$(5S) was found.

We fully reconstruct six modes
$B_s^0 \to D_s^- \pi^+$, $B_s^0 \to D_s^{*-} \pi^+$,
$B_s^0 \to D_s^- \rho^+$, $B_s^0 \to D_s^{*-} \rho^+$,
$B_s^0 \to J/\psi \phi$ and $B_s^0 \to J/\psi \eta$,
which have large reconstruction efficiencies and
are mediated by unsuppressed $b \to c$ tree diagrams.
Charge-conjugate modes are implicitly included everywhere in this paper.
To improve the statistical significance of the $B_s^0$ signal,
these six modes are combined; the masses of the $B_s^0$ and $B_s^*$
mesons are determined from a common signal fit.

In addition, we search for several rare $B_s^0$ decays:
the penguin annihilation decay $B_s^0 \to \gamma \gamma$,
the electromagnetic $b\to s$ penguin decay $B_s^0 \to \phi \gamma$, and
the hadronic $b\to s$ penguin decay $B_s^0 \to K^+ K^-$.
Although the branching fractions for these decays are
expected to be too small to be observed with this dataset,
we can obtain useful upper limits.
To date, only upper limits for the decays 
$B_s^0 \to \gamma \gamma$ \cite{lepgg} and $B_s^0 \to \phi \gamma$ \cite{cdfpg}
have been published.
Within the Standard Model the $B_s^0 \to \gamma \gamma$ decay is expected to 
proceed via a penguin annihilation diagram and to have a branching fraction
in the range $(0.5-1.0) \times 10^{-6}$ \cite{gga,ggb}.
However, this decay is sensitive to some beyond-the-Standard 
Model (BSM) contributions and can be enhanced by one to two orders
of magnitude in some BSM models \cite{ggc,ggd}.
Although current measurements of the process $B \to X_s \gamma$ provide
a more restrictive constraint
for many BSM models, in these models the $B_s^0 \to \gamma \gamma$ process 
is more sensitive.

The decay modes $B_s^0 \to \phi \gamma$ and $B_s^0 \to K^+ K^-$
are also mediated by penguin diagrams; these 
decays are natural processes in which to search 
for BSM physics \cite{kkfl,kka,kkb,kkbu,kkc}.
The decay $B_s^0 \to K^+ K^-$ has
been observed by CDF using a simultaneous multi-channel analysis \cite{cdfkk},
where overlapping signal peaks from the $B_s^0 \to K^+ K^-$, 
$B^0 \to K^+ \pi^-$,
$B^0 \to \pi^+ \pi^-$ and $B_s^0 \to K^- \pi^+$ decay modes were separated
statistically in the fit.
In this analysis the ratio 
$(f_s^T / f_d^T) \times {\cal B}(B_s^0 \to K^+ K^-) /
{\cal B}(B^0 \to K^+ \pi^-) = 0.46 \pm 0.08 \pm 0.07$ was
obtained, where $(f_s^T / f_d^T)$ is the ratio of production
fractions of $B_s^0$ and $B^0$ at Tevatron center-of-mass energy
$\sqrt{s}=1.96\,$TeV.

 We have also searched for the $B_s^0 \to D_s^{(*)+} D_s^{(*)-}$ decay modes.
These decay branching fractions are of special interest \cite{gros,dsds}. 
These modes are expected to be predominantly $CP$ eigenstates
and, because their branching fractions are expected to be large,
they should lead to a sizable lifetime difference between the $CP$-odd and 
$CP$-even $B_s^0$ mesons.
Therefore within the SM framework the relative decay-width difference
$\Delta \Gamma_{B_s^0} / \Gamma_{B_s^0}$ can be obtained from
measurement of the $B_s^0 \to D_s^{(*)+} D_s^{(*)-}$ branching 
fractions. The first observation of the $B_s^0 \to D_s^+ D_s^-$ decay
has recently been published by the CDF collaboration \cite{cdfdsds}.

\section{Belle detector and event selection}

The Belle detector operates at KEKB \cite{kekb},
an asymmetric energy double storage ring designed to collide
\mbox{8 GeV} electrons and 3.5 GeV positrons to produce $\Upsilon$(4S)
mesons with a boost of $\beta \gamma$ = 0.425.
In this analysis we use a data sample of $1.86\,\mathrm{fb}^{-1}$
taken at the $\Upsilon$(5S) energy of \mbox{$\sim$10869 MeV} with
the same boost.
The experimental conditions for data taking at the $\Upsilon$(5S)
were identical to those for $\Upsilon$(4S) or continuum running.

The Belle detector is a general-purpose large-solid-angle magnetic
spectrometer
that consists of a silicon vertex detector,
a central drift chamber (CDC), an array of
aerogel threshold \v{C}erenkov counters (ACC), a barrel-like
arrangement of time-of-flight scintillation
counters (TOF), and an electromagnetic calorimeter
comprised of CsI(Tl) crystals (ECL) 
located inside a superconducting
solenoidal coil with a 1.5~T magnetic field.
An iron flux-return located outside the coil is
instrumented to detect $K^0_L$ mesons and to identify muons (KLM).
The detector is described in detail elsewhere~\cite{belle}.
A GEANT-based detailed simulation of the Belle detector is used
to produce Monte Carlo event samples (MC) and determine efficiencies.

Charged tracks are required to have momenta greater than 100$\,$MeV/$c$.
Kaon and pion mass hypotheses are assigned based on a likelihood
ratio ${\cal L}_{K/\pi} = {\cal L}_K/({\cal L}_K + {\cal L}_{\pi})$, obtained
by combining information from the CDC ($dE/dx$), ACC, and TOF systems.
We require ${\cal L}_{K/\pi} > 0.6$ (${\cal L}_{K/\pi} < 0.6$)
for kaon (pion) candidates~\cite{nakano}.
With these requirements, the identification efficiency for particles
used in this analysis varies from 86$\%$ to 91$\%$ (94$\%$ to 98$\%$)
for kaons (pions). 
A tighter kaon identification requirement ${\cal L}_{K/\pi} > 0.8$
is applied for the $B_s^0 \to K^+ K^-$ decay,
where the pion misidentification background is large.

Electrons are identified combining information from the CDC
(specific ionization $dE/dx$), the ACC, and the ECL
(electromagnetic shower position, shape and energy) \cite{ele}.
Muons are identified by matching tracks to KLM hits and 
by using penetration depth information \cite{muo}.

ECL clusters with a photon-like shape
that are not associated with charged tracks
are accepted as photon candidates.
Primary candidate photons ($\gamma$) that are used to reconstruct the
$B_s^0 \to \phi \gamma$ and $B_s^0 \to \gamma \gamma$ decays are required to
have proper bunch-crossing timing and to lie within the acceptance of
the ECL barrel ($33^{\circ} < \theta_{\gamma} < 128^{\circ}$).
To reduce the background from high-energy $\pi^0$ decays where
the two daughter photons have merged into a single cluster in the calorimeter,
the ECL energy deposition in a group of 3$\times$3 cells is
required to exceed 95\% of that in the group of 5$\times$5 cells 
around the maximum energy cell.
The main background sources of high energy
photons are $\pi^0 \to \gamma \gamma$ and
$\eta \to \gamma \gamma$ decays. To reduce these backgrounds,
restrictions are imposed on the invariant mass of the candidate primary
photon and any other photon ($\gamma^{\,\prime}$) in the event.
The primary photon is rejected if
120$\,$MeV/$c^2\, < M(\gamma\gamma^{\,\prime}) < 145\,$MeV/$c^2$ and
$E_{\gamma^{\,\prime}}\,>\,30\,$MeV, or if
510$\,$MeV/$c^2\, < M(\gamma\gamma^{\,\prime}) < 570\,$MeV/$c^2$ and
$E_{\gamma^{\,\prime}}\,>\,200\,$MeV.

Neutral pion candidates are formed from pairs of photons,
each with energy greater than 150\,MeV; the photons must have
an invariant mass within
$\pm 15\,$MeV/$c^2$ of the nominal $\pi^0$ mass
(i.e.\ \mbox{$\sim3\sigma$},
where \mbox{$\sigma \sim 5\,$MeV/$c^2$} is the $\pi^0$ mass resolution).
A mass-constrained kinematic fit is performed
on the $\pi^0$ candidates to improve their energy resolution.
We reconstruct $\eta$ mesons only in the clean $\eta \to \gamma \gamma$ mode,
requiring an invariant mass within 
$\pm 20\,$MeV/$c^2$ ($\sim$\,2$\sigma$) of the nominal $\eta$ mass
and photon energies larger than 50\,MeV.
$K^0_S$ candidates are formed from $\pi^+\pi^-$ pairs
with an invariant mass within $\pm 10\,$MeV/$c^2$ ($\sim$\,3$\sigma$)
of the nominal $K^0_S$ mass and having a common vertex
displaced from the interaction point by more than 0.1\,cm in the
plane perpendicular to the beam direction.

The invariant mass for $K^{*0} \to K^+ \pi^-$ candidates is required to
be within
$\pm 50\,$MeV/$c^2$ of the nominal $K^{*0}$ mass; those of
$\phi \to K^+ K^-$ candidates,
within $\pm 12\,$MeV/$c^2$ of the $\phi$ mass.
A $\pm 100\,$MeV/$c^2$ mass window is used to select
$\rho^+ \to \pi^+ \pi^0$ candidates.
$D_s^-$ mesons are reconstructed in the $\phi \pi^-$, $K^{*0} K^-$
and $K_S^0 K^-$ decay channels; 
all candidates must have a mass within $\pm 12\,$MeV/$c^2$ ($\sim\,2.5\sigma$)
of the nominal $D_s^-$ mass.
The $D_s^-$ helicity angle distributions are expected to be proportional to
\mbox{cos$^2 \theta_{\rm hel}^{D_s}$} for pseudoscalar-vector final states;
thus a $|$cos $\theta_{\rm hel}^{D_s}| > 0.25$ requirement is applied
for the $D_s^- \to \phi \pi^-$ and $D_s^- \to K^{*0} K^-$ decays.
The helicity angle $\theta_{\rm hel}^{D_s}$ is defined
as the angle between the directions of the $K^-$ and $D_s^-$ momenta
in the $\phi$ rest frame (or the directions of the $\pi^-$ and $D_s^-$ momenta
in the $K^{*0}$ rest frame in the case of $K^{*0} K^-$ decay).

$D_s^{*-}$ candidates are reconstructed in the $D_s^{*-} \to D_s^- \gamma$
mode; the measured $D_s^{*-}$ and $D_s^-$ mass difference is required 
to be within $\pm 10\,$MeV/$c^2$ of its nominal value.
The invariant mass of candidate $J/\psi$ mesons is
required to satisfy
$| M(\mu^+ \mu^-) - m_{J/\psi} | < 30\,$MeV/$c^2$ for the muon decay mode
and satisfy
$-100\,$MeV/$c^2\, < M(e^+ e^-) - m_{J/\psi} < 30\,$MeV/$c^2$ for 
the electron decay mode,
where $m_{J/\psi}$ is the nominal $J/\psi$ mass.

$B_s^0$ decays are reconstructed in the following final states:
$D_s^- \pi^+$, $D_s^- \rho^+$, $D_s^{*-} \pi^+$, $D_s^{*-} \rho^+$, 
$J/\psi \phi$, $J/\psi \eta$, $D_s^{(*)+} D_s^{(*)-}$,
$K^+ K^-$, $\phi \gamma$, and $\gamma \gamma$.
The signals can be observed using two variables:
the energy difference $\Delta E\,=\,E^{CM}_{B_s^0}-E^{\rm CM}_{\rm beam}$
and the beam-energy-constrained mass
$M_{\rm bc} = \sqrt{(E^{\rm CM}_{\rm beam})^2\,-\,(p^{\rm CM}_{B_s^0})^2}$,
where $E^{\rm CM}_{B_s^0}$ and $p^{\rm CM}_{B_s^0}$ are the energy and momentum
of the $B_s^0$ candidate in the $e^+ e^-$ center-of-mass (CM) system,
and $E^{\rm CM}_{\rm beam}$ is the CM beam energy.
The $B_s^0$ mesons can be produced in $e^+ e^-$ collisions
at the $\Upsilon$(5S) energy
via intermediate $B_s^* \bar{B}_s^*$, $B_s^* \bar{B}_s^0$,
$B_s^0 \bar{B}_s^*$ and $B_s^0 \bar{B}_s^0$ channels,
with $B_s^* \to B_s^0 \gamma$.
These intermediate channels can be distinguished kinematically
in the $M_{\rm bc}$ and $\Delta E$ plane, where 
three well-separated $B_s^0$ signal regions
can be defined corresponding to the cases where both, only one, or neither
of the $B_s^0$ mesons originate from a $B_s^*$ decay.
The events obtained from MC simulation of the 
$B_s^0 \to D_s^- \pi^+$ decay are shown in Fig. 1 for the
intermediate $\Upsilon$(5S) decay channels 
$B_s^* \bar{B}_s^*$, $B_s^* \bar{B}_s^0$,
$B_s^0 \bar{B}_s^*$ and $B_s^0 \bar{B}_s^0$.
The signal regions are defined as ellipses
corresponding to \mbox{$\pm$(2.0--2.5)$\sigma$} (i.e. (95-98)\% acceptance) 
resolution intervals in $M_{\rm bc}$ and $\Delta E$.
The signal events from the different intermediate channels are
well separated in the $M_{\rm bc}$ and $\Delta E$ plane.
MC simulation shows that the separation between the channels
in the $M_{\rm bc}$ projection
is $\sim3\,\sigma$ or better for all studied $B_s^0$ decays.
Elliptical regions do not 
describe well the signal shape in the case of $B_s^0$ decays 
to the final states with photons or electrons, because
the radiative energy losses result in a long tail on the left side of
the signal $\Delta E$ distribution. In such decay modes the 
acceptance of the elliptical signal regions
decreases to $(70-80)\,\%$.
A MC simulation indicates that 
the correlation between the $M_{\rm bc}$ and $\Delta E$ variables 
is small and can be neglected in this analysis.
The numbers of events inside and outside these elliptical regions
can be used to estimate the number of $B_s^0$ signal
and background events.

\begin{figure}[h!]
\vspace{-0.1cm}
\epsfig{file=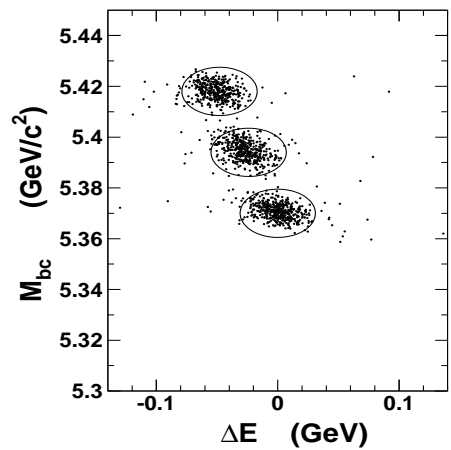,width=6.0cm,height=6.0cm}
\vspace{-0.1cm}
\caption{The $M_{\rm bc}$ and $\Delta E$ scatter plot
for the $B_s^0 \to D_s^- \pi^+$ decay obtained from
the MC simulation. The ellipses show the signal regions for the 
intermediate $B_s^* \bar{B}_s^*$ (top elliptical region), 
$B_s^* \bar{B}_s^0$ and $B_s^0 \bar{B}_s^*$ (middle elliptical region),
and $B_s^0 \bar{B}_s^0$ (bottom elliptical region) channels.}
\label{figamc}
\end{figure}

After all selections the dominant background is from
$e^+e^- \rightarrow q \bar{q}$ continuum events ($q = u,d,s,$ or $c$).
Topologically, $B_s^0$ events are expected to be spherical,
whereas continuum events are expected to be jet-like.
To suppress continuum background, we apply topological cuts.
These were optimized using MC to model the signal and data
outside the $B_s^0$ signal regions to estimate background.
The ratio of the second to the zeroth Fox-Wolfram moments \cite{fox}
is required to be less than 0.3 for the high background
$D_s^{(*)-} \pi^+$,
$D_s^{(*)-} \rho^+$ and $K^+ K^-$ final states,
less than 0.5 for the $\gamma \gamma$ final state
(to increase the signal efficiency of such non-spherical $B_s^0$ decays)
and less than 0.4 for all the other final states.
To suppress continuum further, the angle $\theta^*_{\rm thr}$ in the CM
between the thrust axis of the particles forming the $B_s^0$ candidate
and the thrust axis of all other particles in the event is used.
We require $|{\rm cos}\,\theta^*_{\rm thr}| < 0.9$ for the 
low background final states
with a $J/\psi$, $|{\rm cos}\,\theta^*_{\rm thr}| < 0.7$ for
the $D_s^{(*)-} \rho^+$ final states,
$|{\rm cos}\,\theta^*_{\rm thr}| < 0.6$ for $B_s^0$ events reconstructed
using the $D_s^- \to K^{*0} K^-$ decay mode,
$|{\rm cos}\,\theta^*_{\rm thr}| < 0.5$ for the very high
background $K^+ K^-$ final state,
and $|{\rm cos}\,\theta^*_{\rm thr}| < 0.8$ for all the other final states.

More than one $B_s^0$ candidate per event can be selected. Using MC simulation
we find that $B_s^0$ decays to channels with $D_s^-$ or $D_s^{*-}$ 
mesons can
produce incorrect candidates reconstructed in a cross-channel. 
Because the photon from the $D_s^{*-} \to D_s^- \gamma$ decay has
a low energy, this photon can be removed from the $B_s^0$ reconstruction
resulting in the replacement of an original $D_s^{*-}$ by its 
daughter $D_s^-$ or, conversely, an original $D_s^-$ meson can be combined
with a random photon to produce a false $D_s^{*-}$ candidate.
For example, every $B_s^0 \to D_s^{*-} \pi^+$ decay will
produce an incorrect $B_s^0 \to D_s^- \pi^+$ candidate and
$\sim$37$\,\%$ of $B_s^0 \to D_s^- \pi^+$ decays will produce
incorrect $B_s^0 \to D_s^{*-} \pi^+$ candidates.
Moreover, multiple candidates can be reconstructed 
in the $B_s^0 \to D_s^{*-} \pi^+$ decay mode if the original photon
from the $D_s^{*-}$ decay is replaced by a random photon that satisfies
the $D_s^{*-}$ mass window requirement.
The $M_{\rm bc}$ distribution of incorrectly reconstructed 
$B_s^0$ candidates has the same 
central value as the original signal, but the width is slightly larger.
However, the $\Delta E$ distribution of these incorrectly reconstructed
$B_s^0$ candidates is $(200-300)\,$MeV wide and shifted to
negative values if the correct photon is lost and to
positive values when a random photon is added.
 
We checked the effects of incorrectly reconstructed $B_s^0$ candidates 
on the results of the measurements reported in this paper. Because of the
large spread in the $\Delta E$ distribution of the incorrectly reconstructed
candidates and the 
low statistics used in this analysis, these effects are found to 
be small and are neglected; the corresponding uncertainties 
are included in the systematic error.
We also checked other sources of multiple candidates
in all studied decay modes and found that these effects can
be neglected in the $M_{\rm bc}$ and $\Delta E$ measurements
presented below.
It should be noted that the MC efficiency calculations also include 
multiple candidates and, therefore, a corresponding correction for this 
effect is applied.

\section{Study of \boldmath{$B_{\tiny s}^0 \to D_{\tiny s}^{(*)-} \pi^+$},
\boldmath{$B_{\tiny s}^0 \to D_{\tiny s}^{(*)-} \rho^+$},
\boldmath{$B_{\tiny s}^0 \to J/\psi \phi$}
and \boldmath{$B_{\tiny s}^0 \to J/\psi \eta$} decays}

\begin{figure*}[t!]
\vspace{-0.1cm}
\epsfig{file=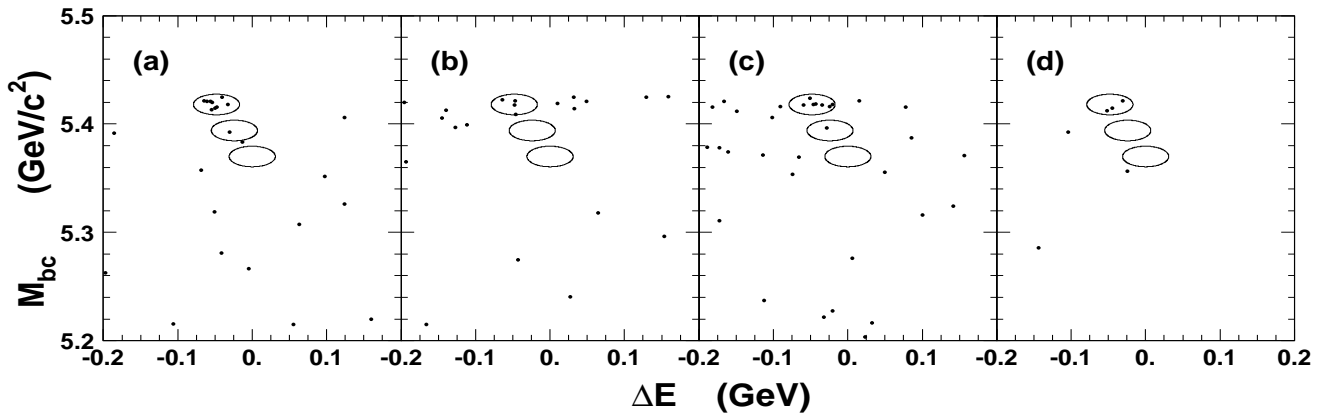,width=17.5cm,height=6.3cm}
\vspace{-0.4cm}
\caption{The $M_{\rm bc}$ and $\Delta E$ scatter plots
for the $B_s^0 \to D_s^- \pi^+$ (a),
$B_s^0 \to D_s^{*-} \pi^+$ (b), 
$B_s^0 \to D_s^{(*)-} \rho^+$ (c) and
$B_s^0 \to J/\psi \phi$ and $B_s^0 \to J/\psi \eta$ (d) decay modes.}
\label{figall}
\end{figure*}

The $M_{\rm bc}$ versus $\Delta E$ distribution 
for the $B_s^0 \to D_s^- \pi^+$ candidates is shown in \mbox{Fig.\ 2a}.
Nine events are observed within the elliptical signal
region corresponding to the $B_s^* \bar{B}_s^*$ pair production channel.
Only one event is observed in the signal region for the
$B_s^* \bar{B}_s^0 + B_s^0 \bar{B}_s^*$ channels, and no events are observed for
the $B_s^0 \bar{B}_s^0$ channel.
Background outside the signal regions is small and 
corresponds to $\sim$0.1 events in each of the three signal regions.
The total number of $b\bar{b}$ events in the sample and the fraction of
$B_s^{(*)} \bar{B}_s^{(*)}$ events among all $b\bar{b}$ events
at the $\Upsilon$(5S) have been determined in \cite{bela} to be 
${\it N}^{b\bar{b}}_{\rm 5S} = (5.61 \pm 0.03_{\rm stat} \pm 0.29_{\rm syst}) \times 10^5$ and \mbox{$f_s = (18.0 \pm 1.3 \pm 3.2)\%$}, 
respectively. We assume that 100$\%$ of $B_s^*$ mesons
decay to the ground state $B_s^0$.
From the 10 observed events, the background estimate of 0.3 events, and the
full reconstruction efficiency of $(0.71 \pm 0.10)\%$ (intermediate branching
fractions are included), we measure the branching fraction 
\mbox{${\cal B}(B_s^0 \rightarrow D_s^- \pi^+)\, =$}
$(0.68 \pm 0.22 \pm 0.16)\%$.
The systematic error includes the ${\rm N}^{b\bar{b}}_{\rm 5S}$ and 
$f_s$ uncertainties and the uncertainty of \mbox{$\sim 14\%$} in the
reconstruction efficiency, which is dominated by the uncertainty in
the value of ${\cal B} (D_s^-\,\to\,\phi \pi^-)$.
This branching fraction is consistent with the value
\mbox{${\cal B}(B_s^0 \rightarrow D_s^- \pi^+)\, =$}
$(0.38 \pm 0.05 \pm 0.14)\%$ derived from a CDF
measurement of ${\cal B}(B_s^0 \rightarrow D_s^- \pi^+) / 
{\cal B}(B^0 \rightarrow D^- \pi^+)$ \cite{cdfbf}
using the 2006 PDG values of the 
$B^0 \rightarrow D^- \pi^+$ and 
$D_s^- \rightarrow \phi \pi^-$ branching fractions \cite{pdg}.

$M_{\rm bc}$ and $\Delta E$ scatterplots
are also obtained for the $B_s^0 \to D_s^{*-} \pi^+$ (Fig.\ 2b) 
and $B_s^0 \to D_s^{(*)-} \rho^+$ (Fig.\ 2c) decay modes.
We observe four $B_s^0 \to D_s^{*-} \pi^+$ candidates and
seven $B_s^0 \to D_s^{(*)-} \rho^+$ candidates in the $B_s^* \bar{B}_s^*$
channel, one $B_s^0 \to D_s^{(*)-} \rho^+$ candidate in the
$B_s^* \bar{B}_s^0 + B_s^0 \bar{B}_s^*$ channel, and no candidates
in the $B_s^0 \bar{B}_s^0$ channel.

The scatterplot in $M_{\rm bc}$ and $\Delta E$ for the 
$B_s^0 \to J/\psi \phi$ and $B_s^0 \to J/\psi \eta$
decays is shown in Fig.\ 2d. Two candidates
are reconstructed in the $B_s^0 \to J/\psi \phi$ mode and one
candidate is reconstructed in the $B_s^0 \to J/\psi \eta$ mode.
One of the observed $B_s^0 \to J/\psi \phi$ candidates
is reconstructed in the $J/\psi \to \mu^+ \mu^-$ mode and one
in the $J/\psi \to e^+ e^-$ mode. As a cross-check, 
the branching fraction
\mbox{${\cal B}(B_s^0 \to J/\psi \phi)\, =$}
$(0.9 \pm 0.6 \pm 0.2) \times 10^{-3}$ is obtained for
these two candidates,
which agrees with the CDF measurement \cite{cdfjp} within the large errors.
The numbers of $B_s^0$ candidates reconstructed in the 
$D_s^- \pi^+$, $D_s^- \rho^+$, $D_s^{*-} \pi^+$, 
$D_s^{*-} \rho^+$, $J/\psi \phi$ and $J/\psi \eta$ decay modes
and lying in the signal region corresponding to
the $B_s^* \bar{B}_s^*$ channel are listed in Table~I. In addition,
the numbers
of events reconstructed in the three $D_s^-$ decay modes are shown separately. 

\renewcommand{\arraystretch}{1.3}
\begin{table}[h!]
\caption{The number of the $B_s^0$ candidates located within the
elliptical signal region corresponding to 
the $B_s^* \bar{B}_s^*$ channel.
The events reconstructed in the $D_s^- \to \phi \pi^-$, 
$D_s^- \to K^{*0} K^-$ and $D_s^- \to K_S^0 K^-$ decay modes
are listed separately.}
\vspace{0.2cm}
\label{tab:bfr2}
\begin{tabular}
{@{\hspace{0.1cm}}l@{\hspace{0.1cm}} @{\hspace{0.1cm}}c@{\hspace{0.1cm}} @{\hspace{0.1cm}}c@{\hspace{0.1cm}} @{\hspace{0.3cm}}c@{\hspace{0.1cm}} @{\hspace{0.1cm}}c@{\hspace{0.1cm}} @{\hspace{0.3cm}}c@{\hspace{0.1cm}} }
\hline \hline
Decay mode & $D_s^- \to$ & $\phi \pi^-$ & $K^{*0} K^-$ & $K_S^0 K^-$ &  Sum \\
\hline
$B_s^0 \to D_s^- \pi^+$ & & 4  & 2 & 3 & 9 \\
$B_s^0 \to D_s^{*-} \pi^+$ & & 2  & 1 & 1 & 4 \\
$B_s^0 \to D_s^- \rho^+$ & & 2  & 1 & 0 & 3 \\
$B_s^0 \to D_s^{*-} \rho^+$ & & 2  & 2 & 0 & 4 \\
$B_s^0 \to J/\psi \phi$ & & & & & 2 \\
$B_s^0 \to J/\psi \eta$ & & & & & 1 \\
\hline \hline
\end{tabular}
\vspace{-0.1cm}
\end{table}

Although the $M_{\rm bc}$ and $\Delta E$ signal resolutions are slightly different
for different decay modes, for simplicity in Fig.\ 2 the same size elliptical 
signal regions are shown for all modes.
Due to low statistics, a small variation of signal shape for different
$B_s^0$ decays can be neglected in the $M_{\rm bc}$ 
and $\Delta E$ measurements discussed below. It is, however,
included in the systematic uncertainties.

The six $B_s^0$ modes shown in Fig.\ 2 are combined
to increase the statistical significance of the $B_s^0$
signal. Distributions in $\Delta E$ are obtained
separately for events from three $M_{\rm bc}$ intervals,
$5.408\,$GeV/$c^2\, < M_{\rm bc} < 5.429\,$GeV/$c^2$ (Fig.\ 3a),
$5.384\,$GeV/$c^2\, < M_{\rm bc} < 5.405\,$GeV/$c^2$ (Fig.\ 3b) and
$5.360\,$GeV/$c^2\, < M_{\rm bc} < 5.380\,$GeV/$c^2$ (Fig.\ 3c), 
corresponding to 
$B_s^0$ production proceeding through the $B_s^* \bar{B}_s^*$, 
$B_s^* \bar{B}_s^0 + B_s^0 \bar{B}_s^*$ or $B_s^0 \bar{B}_s^0$
channels, respectively.

Each of these three distributions is fitted with the sum of a Gaussian
to describe the signal and a linear function to describe the background.
In the $B_s^* \bar{B}_s^*$ channel \mbox{(Fig.\ 3a)}, the width and the peak 
position are allowed to float, and their values
$\sigma_{\Delta E} = (10.2 \pm 1.9)\,$MeV and 
$\langle \Delta E \rangle = (-47.6 \pm 2.6)\,$MeV, respectively, are obtained from the fit.
The width agrees with the value of $\sim 12\,$MeV
obtained from a MC simulation of the dominant 
$B_s^0 \rightarrow D_s^+ \pi^-$ decay channel.
Due to low statistics in the other two distributions, 
the peak positions and widths are fixed. The widths
are taken from MC simulations.
The peak position is fixed to zero for the $B_s^0 \bar{B}_s^0$ channel
and that for the $B_s^* \bar{B}_s^0 + B_s^0 \bar{B}_s^*$ channel
is fixed to $-23.8$\,MeV, which is
half of the value obtained for the $\langle \Delta E \rangle$ peak position 
in the $B_s^* \bar{B}_s^*$ channel. 
The fits yield $20.3 \pm 4.8$ events and
$1.5 \pm 2.0$ events for the $B_s^* \bar{B}_s^*$ and 
$B_s^* \bar{B}_s^0 + B_s^0 \bar{B}_s^*$ channels, respectively;
no events are observed in the $B_s^0 \bar{B}_s^0$ channel.
From these numbers and approximately equal $B_s^0$ reconstruction
efficiency in these three channels found in MC simulation, we obtain
the ratio $\sigma (e^+ e^- \to B_s^* \bar{B}_s^*) /
\sigma (e^+ e^- \to B_s^{(*)} \bar{B}_s^{(*)}) = (93 ^{+7}_{-9} \pm 1)\%$
at the $\Upsilon$(5S) energy. The first uncertainty is statistical and 
the second uncertainty is systematic, dominated by uncertainties in the
fit procedure.
Potential models predict the fraction of $B_s^* \bar{B}_s^*$
production to be around 70$\%$ \cite{pota,potb,potc}.

\begin{figure}[t!]
\vspace{-0.1cm}
\begin{center}
\epsfig{file=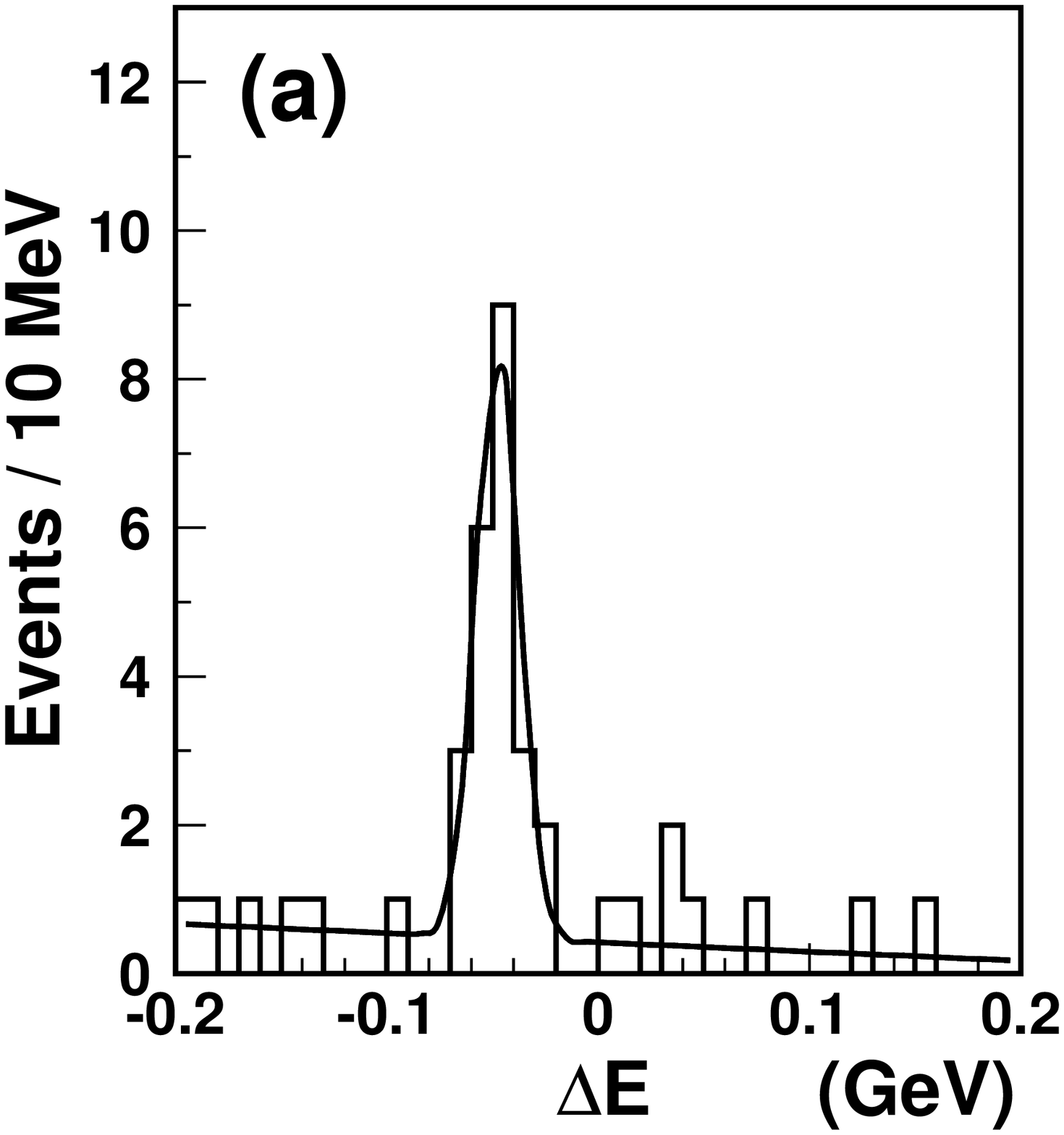,width=4.2cm,height=4.2cm}\epsfig{file=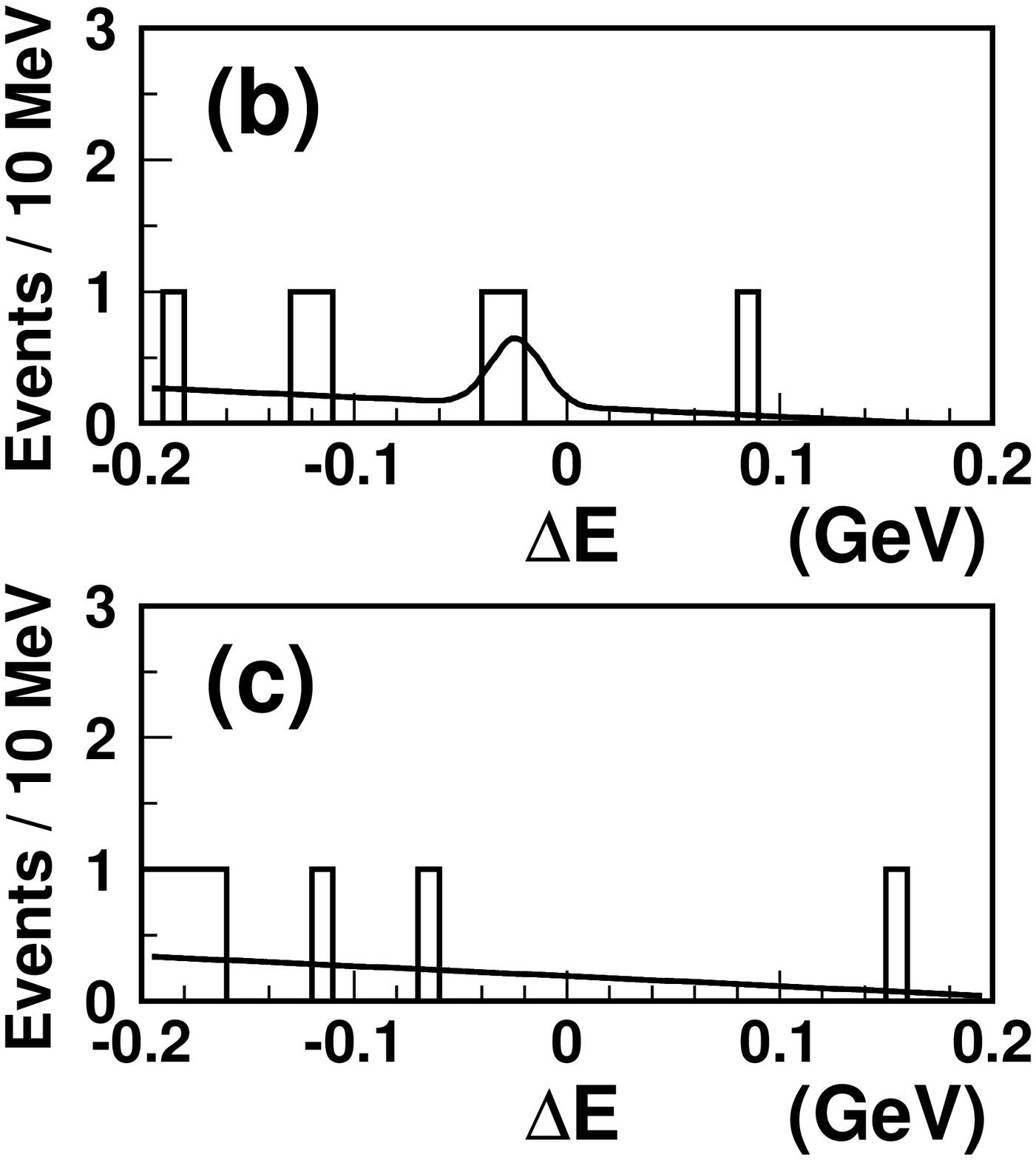,width=4.2cm,height=4.2cm}
\end{center}
\vspace{-0.3cm}
\caption{The $\Delta E$ distributions
for $B_s^0$ candidates with
(a) $5.408~$GeV/$c^2 < M_{\rm bc} < 5.429\,$GeV/$c^2$,
(b) $5.384~$GeV/$c^2 < M_{\rm bc} < 5.405\,$GeV/$c^2$ and
(c) $5.360~$GeV/$c^2 < M_{\rm bc} < 5.380\,$GeV/$c^2$, 
corresponding to
$B_s^0$ production through the $B_s^* \bar{B}_s^*$,
$B_s^* \bar{B}_s^0 + B_s^0 \bar{B}_s^*$ and $B_s^0 \bar{B}_s^0$
channels, respectively. 
Curves represent the results of the fits described in the text.}
\label{figala}
\end{figure}

The $B_s^*$ mass can be extracted from fit to
the $M_{\rm bc}$ distribution of the observed events
in the $B_s^* \bar{B}_s^*$ channel. 
In this channel the $M_{\rm bc}$ variable, calculated from the formula
$M_{\rm bc}=\sqrt{(E^{\rm CM}_{\rm beam})^2\,-\,(p^{\rm CM}_{B_s^0})^2}$,
is equal, to a good approximation, to the mass of $B_s^*$ meson.
This follows from the fact
that the difference between the $B_s^0$ and $B_s^*$ momenta 
is statistically unbiased from zero and
is smaller than the experimental resolution in $B_s^0$ momentum.
Figure 4 shows the $M_{\rm bc}$ distribution of the candidates in the range
$-80\,$MeV$ < \Delta E < -20\,$MeV, where signal events 
from the $B_s^* \bar{B}_s^*$
production channel are expected. We fit this distribution with
the sum of a Gaussian to
describe the signal and a so-called ARGUS function \cite{argus}
to describe the background. The fit yields a mass value of
$M(B_s^*) = (5418 \pm 1 \pm 3)\,$MeV/$c^2$. The large systematic error is
dominated by the uncertainty in the collider beam energy 
calibration resulting in
a $e^+ e^-$ CM beam energy uncertainty of $\sim$3$\,$MeV.
The uncertainty of the method used to determine the $M(B_s^*)$ mass is 
estimated by MC simulation to be around 0.5$\,$MeV/$c^2$.
The uncertainty in the particle momenta measurements translated to the
$M(B_s^*)$ mass uncertainty is also around 0.5$\,$MeV/$c^2$. 
The observed width of the $B_s^*$ signal is $(3.6 \pm 0.6)\,$MeV/$c^2$ and 
agrees with the value obtained from the MC simulation, which assumes 
zero natural width and is dominated by the KEKB energy spread.
The obtained $B_s^*$ mass is $1.8\,\sigma$ higher than
the value measured recently by CLEO \cite{cleobm}, 
$M(B_s^*) = (5411.7 \pm 1.6 \pm 0.6)\,$MeV/$c^2$.

\begin{figure}[h!]
\vspace{-0.1cm}
\epsfig{file=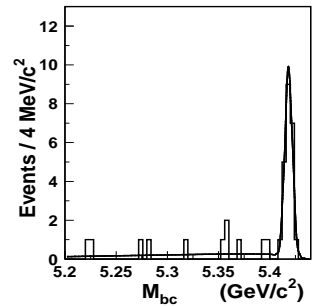,width=4.2cm,height=4.2cm}
\vspace{-0.1cm}
\caption{The $B_s^*$ mass distribution for events
within the $-80~$MeV$ < \Delta E < -20\,$MeV interval, 
where the $B_s^0$ signal from the $B_s^* \bar{B}_s^*$ channel is expected.
Curve represents the result of the fit described in the text.}
\label{figalb}
\end{figure}

Using the measured values of the $B_s^*$ mass and the energy difference
$\langle \Delta E \rangle = (-47.6 \pm 2.6)\,$MeV,
we can calculate the $B_s^0$ mass. 
The value $\langle \Delta E \rangle$ is the mean energy difference
between the $B_s^*$ and $B_s^0$ mesons in the CM system and, 
in a good approximation,
is equal to the mass difference of the $B_s^*$ and $B_s^0$ mesons. 
The photon energy in the $B_s^* \to B_s^0 \gamma$ decay is a constant 
in the $B_s^*$ rest frame, and the smearing due 
to the Lorentz transformation from the $B_s^*$ rest frame to the CM rest 
frame is small compared with the central value of the photon energy. 
Finally we obtain a mass value of
$M(B_s^0) = (5370 \pm 1 \pm 3)\,$MeV/$c^2$.
The second uncertainty in the $B_s^0$ mass value is the systematic
uncertainty dominated by the statistical uncertainty on 
the $\langle \Delta E \rangle$ measurement,
which will improve once more statistics become available.
The uncertainty due to the collider beam energy calibration 
almost lineary affects both the
$M(B_s^*)$ and $\langle \Delta E \rangle$ values and
nearly cancels in the $M(B_s^0)$ mass calculations.
Other systematic uncertainties affecting the $B_s^0$ mass
are similar to those in the $B_s^*$ mass measurement and are small.
The obtained $B_s^0$ mass agrees well with the PDG value,
$M(B_s^0) = (5369.6 \pm 2.4)\,$MeV/$c^2$ \cite{pdg}, and the most 
recent CDF measurement,
$M(B_s^0) = (5366.01 \pm 0.73 \pm 0.33)\,$MeV/$c^2$ \cite{cdfmb}.

\section{Search for \boldmath{$B_{\tiny s}^0 \to \gamma \gamma$}, 
\boldmath{$B_{\tiny s}^0 \to \phi \gamma$}, \boldmath{$B_{\tiny s}^0 \to K^+ K^-$},
and \boldmath{$B_{\tiny s}^0 \to D_{\tiny s}^{(*)+} D_{\tiny s}^{(*)-}$} decays}

Distributions in $M_{\rm bc}$ and $\Delta E$ are also obtained
for the reconstructed 
$B_s^0 \to \gamma \gamma$ (Fig.\ 5a), $B_s^0 \to \phi \gamma$ (Fig.\ 5b), 
$B_s^0 \to K^+ K^-$ (Fig.\ 5c) and 
$B_s^0 \to D_s^{(*)+} D_s^{(*)-}$ (Fig.\ 5d) candidates.
Only the $B_s^0$ signal regions corresponding to the dominant
$B_s^* \bar{B}_s^*$ channel are considered for the searches reported here.
These regions are wider for the $B_s^0 \to \phi \gamma$ and
$B_s^0 \to \gamma \gamma$ decays, where energy losses
due to photon radiation lead to a large tail at lower values of 
$\Delta E$. The signal region for the $B_s^0 \to D_s^{(*)+} D_s^{(*)-}$
search is slightly smaller, because of the kinematics of the decay 
to two heavy particles.
The shapes of the signal regions for these decays
are optimized from the MC simulation.

\begin{figure*}[t!]
\vspace{-0.1cm}
\epsfig{file=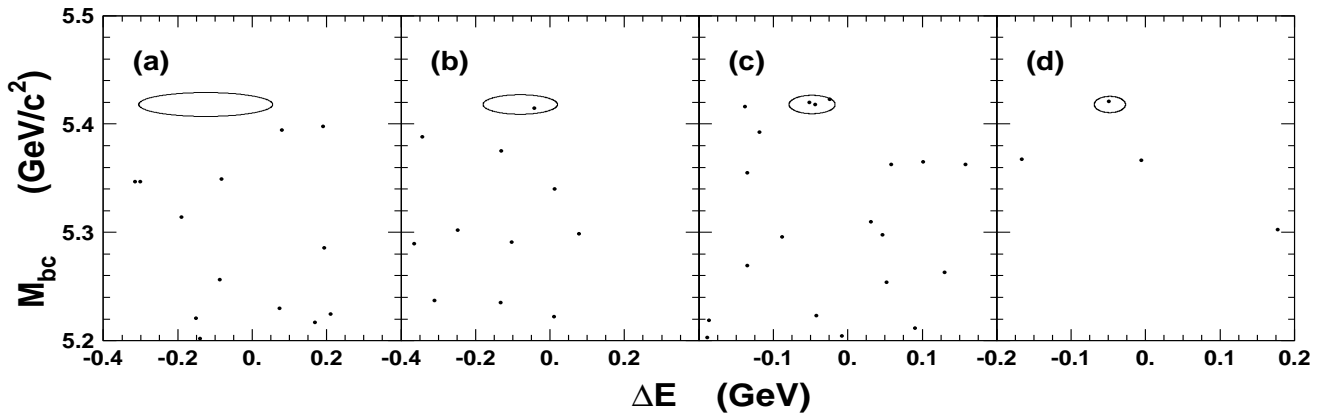,width=17.5cm,height=6.3cm}
\vspace{-0.4cm}
\caption{The scatter plots in $M_{\rm bc}$ and $\Delta E$
for the $B_s^0 \to \gamma \gamma$ (a), $B_s^0 \to \phi \gamma$ (b),
$B_s^0 \to K^+ K^-$ (c) and
$B_s^0 \to D_s^{(*)+} D_s^{(*)-}$ (d) decay modes.
In the latter case, the signal event is reconstructed in the 
$B_s^0 \to D_s^{*+} D_s^-$ decay mode, while the three background events 
are reconstructed in the $B_s^0 \to D_s^+ D_s^-$ decay mode.
The ellipses indicate the $B_s^0$ signal 
regions for the $B_s^* \bar{B}_s^*$ channel.}
\label{figalc}
\end{figure*}

To avoid multiple $D_s^*$ and $D_s$ cross-channel candidates,
only one candidate per event is selected in the 
$B_s^0 \to D_s^{(*)+} D_s^{(*)-}$ analysis, where the number of multiple
candidates can be rather large. The candidate with 
the $M_{\rm bc}$ value closest to the nominal $M(B_s^*)$ value
measured above is chosen. 
No significant signals are observed in any of the distributions
shown in Fig.\ 5.
However one $B_s^0 \to \phi \gamma$ event,
two $B_s^0 \to K^+ K^-$ events and
one $B_s^0 \to D_s^{*+} D_s^-$ event
lie within the signal regions, whereas
backgrounds outside the signal regions are not
large.
The numbers of events within the signal regions, the estimated
background contributions, the efficiencies, and the upper limits
for the corresponding $B_s^0$ branching fractions are listed in Table~II. 
For comparison, the previously published upper limits and
branching fractions are also shown. 
The numbers of events and the upper limits are obtained using
only the $B_s$ signal region corresponding to the
$B_s^* \bar{B}_s^*$ channel.
The upper limits are obtained using the Feldman-Cousins method \cite{feco},
and a small correction due to systematic uncertainties is applied.
The efficiencies are determined from the MC simulation.
The number of initial $B_s^* \bar{B}_s^*$ pairs is obtained 
by multiplying the number of $B_s^{(*)} \bar{B}_s^{(*)}$ pairs
measured in the inclusive analysis \cite{bela} by the production ratio
of $B_s^* \bar{B}_s^*$ pairs to all $B_s^{(*)} \bar{B}_s^{(*)}$ pairs
obtained in this analysis.
We calculated the previous $B_s^0 \to D_s^+ D_s^-$ branching fraction listed
in Table~II using the measurement
${\cal B}(B_s^0 \to D_s^+ D_s^-) / {\cal B}(B^0 \to D_s^+ D^-) = 
1.67 \pm 0.41 \pm 0.47$ from CDF \cite{cdfdsds}
and the ${\cal B}(B^0 \to D_s^+ D^-)$ value from the PDG \cite{pdg}.

\renewcommand{\arraystretch}{1.3}
\begin{table*}[t!]
\caption{The number of events in the signal region (Yield),
the estimated background contribution (Bkg.), the efficiencies (Eff.),
the 90$\%$ C.L. upper limits derived in this 
analysis (Belle {\rm upper limit})
and previously published upper limits
or branching fractions \mbox{(Previous\ {\rm UL/BF})} 
for the $B_s^0 \to \gamma \gamma$, $B_s^0 \to \phi \gamma$,
$B_s^0 \to K^+ K^-$ and $B_s^0 \to D_s^{(*)+} D_s^{(*)-}$ decay modes.}
\vspace{0.2cm}
\label{tab:bfr}
\begin{tabular}
{@{\hspace{0.3cm}}l@{\hspace{0.3cm}} @{\hspace{0.3cm}}c@{\hspace{0.3cm}} @{\hspace{0.3cm}}c@{\hspace{0.3cm}} @{\hspace{0.3cm}}c@{\hspace{0.3cm}} @{\hspace{0.3cm}}c@{\hspace{0.3cm}} @{\hspace{0.3cm}}c@{\hspace{0.3cm}}}
\hline \hline
Decay mode & Yield & Bkg. & Eff. &  Belle & Previous \\
      & (events) & (events) & ($\%$) & upper limit & UL/BF \\ \hline
$B_s^0 \to \gamma \gamma$ & 0  & 0.5 & 20.0 & $<\,$0.53$\,\times 10^{-4}$ & $<\,$1.48$\,\times 10^{-4}$ \cite{lepgg} \\
$B_s^0 \to \phi \gamma$ & 1 & 0.15 & 5.9  & $<\,$3.9$\,\times 10^{-4}$ & $<\,$1.2$\,\times 10^{-4}$ \cite{cdfpg} \\
$B_s^0 \to K^+ K^-$ & 2 & 0.16 & 9.8 & $<\,$3.1$\,\times 10^{-4}$ & $(3.30 \pm 0.57 \pm 0.67)\,\times 10^{-5}$ \cite{cdfkk} \\
$B_s^0 \to D_s^+ D_s^-$ & 0 & 0.02 & 0.020 & $<\,$6.7$\,\%$ & $(1.09 \pm 0.27 \pm 0.47)\,\%$ \cite{cdfdsds} \\
$B_s^0 \to D_s^{*+} D_s^-$ & 1 & 0.01 & 0.0099 & $<\,$12.1$\,\%$ & - \\
$B_s^0 \to D_s^{*+} D_s^{*-}$ & 0 & $<$0.01 & 0.0052 & $<\,$25.7$\,\%$ & - \\
\hline \hline
\end{tabular}
\vspace{-0.1cm}
\end{table*}

The upper limit obtained for the decay $B_s^0 \to \gamma \gamma$
is about three times smaller than the most restrictive 
published limit \cite{lepgg}.
However, it is still two orders of magnitude 
above SM predictions \cite{gga,ggb}.
The upper limit obtained for $B_s^0 \to \phi \gamma$
is about a factor ten larger than the theoretically 
expected branching fraction \cite{phiga}.
The upper limit obtained for the $B_s^0 \to K^+ K^-$ decay
is an order of magnitude 
larger than the value measured by CDF \cite{cdfkk}.
For SM branching fractions,
statistically significant signals of $\sim$10 events 
can be obtained for the $B_s^0 \to \phi \gamma$ and 
$B_s^0 \to K^+ K^-$ modes
in a $\sim$30\,fb$^{-1}$ dataset on the $\Upsilon$(5S).

The upper limits obtained for
$B_s^0 \to D_s^{(*)+} D_s^{(*)-}$ decays
are of special interest because
the $D_s^{(*)+} D_s^{(*)-}$ states are expected to be
dominantly $CP$ eigenstates.
Assuming that the branching fractions for the $D_s^+ D_s^-$,
$D_s^{*+} D_s^-$, $D_s^+ D_s^{*-}$ and $D_s^{*+} D_s^{*-}$ final states
are each in the range (1--3)$\%$, we expect about 5--10 events in each
of these four channels with statistics of $\sim$30\,fb$^{-1}$.
Within the SM framework such measurements can provide an important 
constraint on the value 
of $\Delta \Gamma_{B_s^0} / \Gamma_{B_s^0}$ \cite{gros,dsds}.

\section{Conclusions}

Several exclusive $B_s^0$ decays are reconstructed using 1.86\,fb$^{-1}$
of data taken at the $\Upsilon$(5S) resonance with the Belle 
detector at the KEKB asymmetric energy $e^+ e^-$ collider.

$B_s^0$ signals are found in six decay modes:
$B_s^0 \to D_s^{(*)-} \pi^+$, 
$B_s^0 \to D_s^{(*)-} \rho^+$, 
$B_s^0 \to J/\psi \phi$ 
and $B_s^0 \to J/\psi \eta$.
The branching fraction
\mbox{${\cal B}(B_s^0 \rightarrow D_s^- \pi^+)\, =$}
\mbox{$(0.68 \pm 0.22 \pm 0.16)\%$} is measured.
Combining the studied six channels, we observe a significant
$B_s^0$ signal and obtain the masses 
$M(B_s^0) = (5370 \pm 1 \pm 3)\,$MeV/$c^2$ and 
$M(B_s^*) = (5418 \pm 1 \pm 3)\,$MeV/$c^2$.
$B_s^0$ production through the 
$B_s^* \bar{B}_s^*$ channel is found to dominate over other
$B_s^{(*)} \bar{B}_s^{(*)}$ channels;
the ratio $\sigma (e^+ e^- \to B_s^* \bar{B}_s^*) /
\sigma (e^+ e^- \to B_s^{(*)} \bar{B}_s^{(*)}) = (93 ^{+7}_{-9} \pm 1)\%$
is measured. These results are in agreement with
CLEO measurements \cite{cleoe}.

We have also searched for
$B_s^0 \to \gamma \gamma$, $B_s^0 \to \phi \gamma$, $B_s^0 \to K^+ K^-$
and $B_s^0 \to D_s^{(*)+} D_s^{(*)-}$ decay modes
and set upper limits on their branching fractions.
The upper limit on $B_s^0 \to \gamma \gamma$ is three times more
restrictive than the best existing limit.
The background levels in these decays are low, indicating that the
sensitivity of future studies of these decays with larger statistics
will not be limited by backgrounds.
We expect that significant signals for $B_s^0 \to K^+ K^-$, 
$B_s^0 \to \phi \gamma$ and $B_s^0 \to D_s^{(*)+} D_s^{(*)-}$ decays
can be observed in $\sim$30\,fb$^{-1}$ of data. With such statistics the 
upper limit for the $B_s^0 \to \gamma \gamma$ decay should provide 
an important constraint on some BSM models.

We thank the KEKB group for the excellent operation of the
accelerator, the KEK cryogenics group for the efficient
operation of the solenoid, and the KEK computer group and
the National Institute of Informatics for valuable computing
and Super-SINET network support. We acknowledge support from
the Ministry of Education, Culture, Sports, Science, and
Technology of Japan and the Japan Society for the Promotion
of Science; the Australian Research Council and the
Australian Department of Education, Science and Training;
the National Science Foundation of China and the Knowledge
Innovation Program of the Chinese Academy of Sciences under
contract No.~10575109 and IHEP-U-503; the Department of
Science and Technology of India; 
the BK21 program of the Ministry of Education of Korea, 
the CHEP SRC program and Basic Research program 
(grant No.~R01-2005-000-10089-0) of the Korea Science and
Engineering Foundation, and the Pure Basic Research Group 
program of the Korea Research Foundation; 
the Polish State Committee for Scientific Research; 
the Ministry of Education and Science of the Russian
Federation and the Russian Federal Agency for Atomic Energy;
the Slovenian Research Agency;  the Swiss
National Science Foundation; the National Science Council
and the Ministry of Education of Taiwan; and the U.S.\
Department of Energy.

\end{document}